\documentclass[final,5p,times,twocolumn]{elsarticle}

\usepackage{amssymb}
\usepackage{amsmath}
\usepackage{amsthm}
\usepackage{graphicx}
\usepackage{geometry}
\usepackage{siunitx}
\usepackage{subcaption}

\usepackage{url}
\usepackage{lineno}

\journal{NIM A}

\begin{document}

\begin{frontmatter}


\title{Evaluation of Performance of the p-stop fabrication process variations in ATLAS18 strip sensors, Pre- and Post-Irradiation}

\author[FZU]{J.~Kozáková,\corref{cor1}} \cortext[cor1]{Corresponding author.}
\author[SCIPP]{A.~Affolder}
\author[SCIPP]{K.~Affolder}
\author[QMUL]{A.~Awais}
\author[CNM]{E.~Bach}
\author[QMUL]{G.~A.~Beck}
\author[IFIC]{P.~Bernabeu}
\author[QMUL]{A.~J.~Bevan}
\author[QMUL]{Z.~Chen}
\author[JSI]{V.~Cindro}
\author[Tor]{B.~Crick}
\author[Car]{J.~Dandoy}
\author[QMUL]{I.~Dawson}
\author[SCIPP]{V.~Fadeyev}
\author[FZU]{P.~Federičová}
\author[CNM]{C.~Fleta}
\author[UJP]{P.~Gallus}
\author[Car]{C.~Jessiman}
\author[SCIPP]{S.~Katznelson}
\author[Car]{J.~Keller}
\author[Car]{C.~T.~Klein}
\author[Car]{T.~Koffas}
\author[FZU]{J.~Kroll}
\author[FZU]{M.~Kůtová}
\author[FZU]{J.~Kvasnička}
\author[IFIC]{C.~Lacasta}
\author[JSI]{I.~Mandić}
\author[AUS]{R.~R.~Marcelo Gregorio}
\author[SCIPP]{F.~Martinez-McKinney}
\author[FZU]{M.~Mikeštíková}
\author[QMUL]{P.~S.~Miyagawa}
\author[SCIPP]{L.~Morelos-Zaragoza}
\author[Tor]{R.~S.~Orr}
\author[SCIPP]{Q.~Paddock}
\author[IFIC]{C.~Solaz}
\author[IFIC]{U.~Soldevila}
\author[Car]{E.~Staats}
\author[FZU]{P.~Tůma}
\author[CNM]{M.~Ullán}
\author[KEK]{Y.~Unno}
\author[Tor]{M.~Vansteenkiste}
\author[Tor]{P.~Wishart}
\author[QMUL]{S.~C.~Zenz}
\author[Car]{Y.~Zhao}


\affiliation[FZU]{organization={Institute of Physics, Czech Academy of Sciences},
           addressline={Na Slovance 2}, city={Prague}, postcode={18200}, country={Czech Republic}}

\affiliation[SCIPP]{organization={Santa Cruz Institute for Particle Physics (SCIPP), University of California}, city={Santa Cruz}, postcode={CA 95064}, country={USA}}

\affiliation[QMUL]{organization={Particle Physics Research Centre, Queen Mary University of London}, addressline={G. O. Jones Building, Mile End Road}, city={London}, postcode={E1 4NS}, country={United Kingdom}}

\affiliation[CNM]{organization={Instituto de Microelectrónica de Barcelona (IMB-CNM)},       addressline={CSIC, Campus UAB-Bellaterra}, city={Barcelona}, postcode={08193}, country={Spain}}

\affiliation[IFIC]{organization={Instituto de Física Corpuscular, IFIC/CSIC-UV},  addressline={Catedrático José Beltrán 2} , postcode={E-46980 Paterna}, city={Valencia},      country={Spain}}

\affiliation[JSI]{organization={Experimental Particle Physics Department, Jožef Stefan Institute}, addressline={Jamova 39}, city={Ljubljana},postcode={SI-1000}, country={Slovenia}}

\affiliation[Tor]{organization={Department of Physics, University of Toronto}, addressline={60 Saint George St.}, city={Toronto}, postcode={M5S1A7}, state={Ontario},       country={Canada}}

\affiliation[Car]{organization={Physics Department, Carleton University}, addressline={1125 Colonel By Drive}, city={Ottawa}, postcode={K1S 5B6}, state={Ontario},   country={Canada}}

\affiliation[UJP]{organization={UJP PRAHA a.s.}, addressline={Nad Kamínkou 1345},          city={Prague - Zbraslav}, postcode={15610}, country={Czech Republic}}

\affiliation[AUS]{organization={Department of Nuclear Physics \& Accelerator Applications, The Australian National University}, city={Canberra}, postcode={ACT 2600}, country={Australia}}

\affiliation[KEK]{organization={Institute of Particle and Nuclear Study, High Energy Accelerator Research Organization (KEK)}, addressline={1-1 Oho},city={Tsukuba},        postcode={305-0801},  state={Ibaraki},  country={Japan}}


\begin{abstract}
The High-Luminosity upgrade of the Large Hadron Collider (HL-LHC) requires the replacement of the ATLAS Inner Detector with a new all-silicon Inner Tracker (ITk). Radiation-hard  $n^+$-in-$p$ micro-strip sensors were developed for use in the ITk.

During quality control (QC) and quality assurance (QA) procedures performed on ATLAS18 production sensors and test chips, ATLAS ITk strip sensor institutes identified instances of low p-stop density in certain production batches. To investigate the origin of these issues and to try some technological improvements, two dedicated sample batches with four variations of the p-stop fabrication process were produced. These process types (1–4) were systematically evaluated to determine their effectiveness and uniformity. In total, 56 full-size sensors, along with numerous miniature sensors and test chips, were tested. A~number of samples were used in the irradiation campaign reaching the maximum expected levels for the HL-LHC operation after 10 years - with neutrons up to $1.6 \times 10^{15}$ $n_{\textrm{eq}}/\textrm{cm}^2$, and with gamma rays from a $^{60}$Co source up to 66 Mrad.
 
The goal of this evaluation program is to assess the impact of of p-stop fabrication process variations on interstrip isolation performance and to provide feedback for future sensor development. This paper describes the applied test procedures and presents the corresponding results, which demonstrate good sensor performance for all investigated p-stop types. The findings confirm the reliability of the evaluated designs and identify specific advantages that can inform further optimization of strip sensor fabrication.

\end{abstract}


\end{frontmatter}



\section{Introduction} \label{sec:intro}

The High-Luminosity Large Hadron Collider (HL-LHC) upgrade, scheduled for 2030, necessitates the complete replacement of the ATLAS Inner Detector with a new, all-silicon Inner Tracker (ITk) to operate effectively in the high-luminosity environment. The ATLAS18 $n^+$-in-$p$ micro-strip silicon sensors developed for the ATLAS ITk must withstand extreme radiation loads, specifically a non-ionizing fluence up to $\Phi_{\textrm{eq}} = 1.6 \times 10^{15}$ 1 MeV $n_{\textrm{eq}}/\mathrm{cm}^2$ and a Total Ionizing Dose (TID) of 66 Mrad (including a safety factor of 1.5) \cite{{bib:TDR-ATLAS}, {bib:ATLAS}}. Under these extreme conditions, maintaining high interstrip isolation is critical. Loss of isolation increases charge sharing and readout noise, which severely compromises the detector's spatial resolution. In the ATLAS18 strip sensors, this isolation is achieved by introducing a dense $p$-type trace around the $n$-type strips.

Despite these requirements, instances of low p-stop density were noted in specific sensor batches during routine \mbox{ATLAS18} Quality Assurance (QA) and Quality Control (QC) production testing and additional special qualification tests \cite{bib:Orr-QA}. To identify the source of this variation and assess manufacturing robustness, Hamamatsu Photonics, K. K. (HPK) \cite{bib:HPK-web} produced two sample batches featuring four distinct types of p-stop process variations. This dedicated test production was initiated to systematically evaluate performance differences among these process variations on interstrip isolation, both before and after irradiation with different particles. The study focuses on the analysis of these special test batches and on providing feedback to HPK. While the results will not be used to modify the production process for the final ITk detector construction, they offer valuable input for further optimization of the manufacturing process.


\section{Samples and Irradiation}

For this study, HPK produced two dedicated sample batches featuring four p-stop process variations for evaluation by the ATLAS ITk strip institutions. Each batch consisted of seven wafers per process variation, resulting in 28 wafers per batch. In total, 56 wafers were fabricated and distributed to the participating ATLAS ITk institutions. Type 1 corresponds to the conventional p-stop implant process used in standard production batches, while types 2–4 represent slightly modified variants relative to the baseline process. The aim of this study is to investigate how these modified p-stop fabrication processes (types 2–4) affect the quality of interstrip isolation, including its uniformity across the wafer. An overview of the investigated sample types is given in Table \ref{tab:types} \footnote{In general, damage caused by ion implantation is annealed and the implants are activated at temperatures around 1000\,$^\circ$C~\cite{ref:eesemi}. The details of the process variations (Types~1--4) are proprietary information of the vendor.}.

\begin{table}[ht]
    \centering
    \begin{tabular}{|c|c|}
    \hline
       \textbf{type}  &  \textbf{process variation} \\ \hline
      1  & Conventional process conditioning (reference) \\
      2 &  Lower temperature than type 1 \\
      3 &  Little lower temperature than type 2   \\
      4 &  Another method conditioning  \\ \hline
    \end{tabular}
    \caption{p-stop variation process types of samples.}
    \label{tab:types}
\end{table}

\begin{figure}[ht]
\centering
\includegraphics[width=0.7\linewidth]{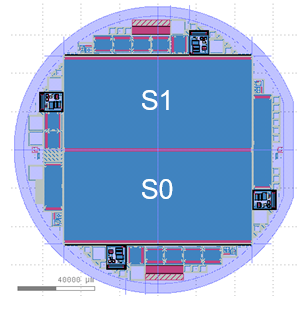}
\caption{Wafer layout of ATLAS18 long strip barrel sensor (LS).}
\label{fig:wafer-LS}
\end{figure}

\begin{figure}[ht]
\centering
\begin{minipage}{0.3\textwidth}
\centering
\includegraphics[width=0.85\linewidth]{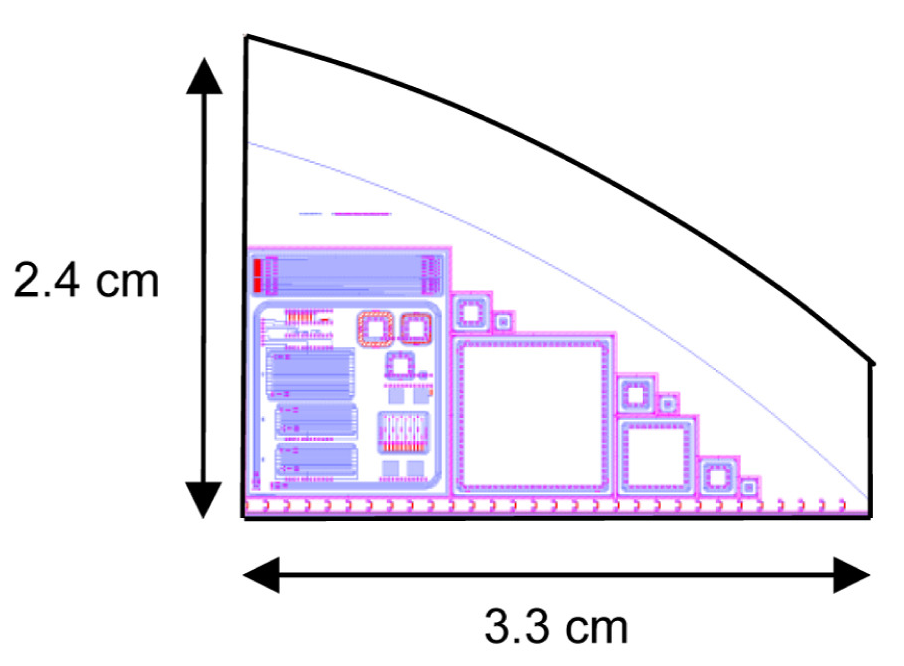}
\subcaption{}
\label{fig:teschip}
\end{minipage}%
\begin{minipage}{0.2\textwidth}
\centering
\includegraphics[width=0.8\linewidth]{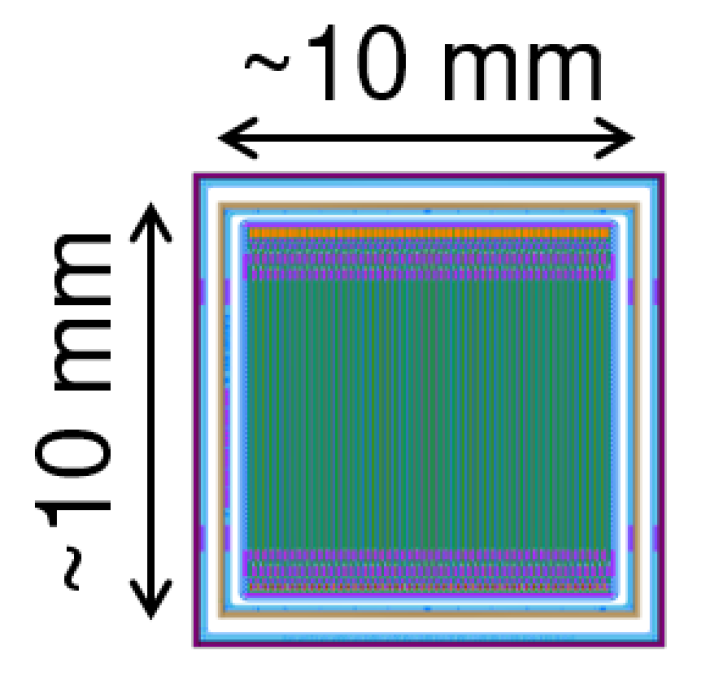}
\subcaption{}
\label{fig:mini}
\end{minipage}
\caption{(a) ATLAS18 QA test chip \cite{bib:Ullan-QA}. (b) ATLAS18 mini sensor \cite{bib:Tejero-thesis}.}
\label{fig:testchip-mini}
\end{figure}

Each production wafer contains a main long strip barrel sensor surrounded by various test structures arranged in four half-moon sections (see Figure \ref{fig:wafer-LS}). The ATLAS18 strip sensors has $n^+$ strips implanted on \mbox{$p$-type} silicon bulk ($n^+$-in-$p$). The aluminum readout strips are AC-coupled to the implant strips, which are biased through polysilicon bias resistors. The tested barrel main sensors have a square geometry 9.8 $\times$ 9.8 cm$^2$, comprising two strip-segment areas (S0 and S1) and 1282 strips in each area\footnote{The first and the last strips are used for electric-field shaping, whereas the remaining 1280 strips are used for readout.}, with a strips length of 48.35~mm and a pitch of 75.5~$\mu$m \cite{bib:Unno-spec}. Figure~\ref{fig:testchip-mini}a shows a detail of one part of half-moon, referred to as the \textit{QA test chip}, which includes a variety of structures dedicated to measurements of specific wafer properties, while Figure~\ref{fig:testchip-mini}b presents a mini sensor. The mini sensors feature a reduced strip length of 8 mm while maintaining the same strip pitch.

\begin{figure}[h!]
    \centering
    \includegraphics[width=6cm]{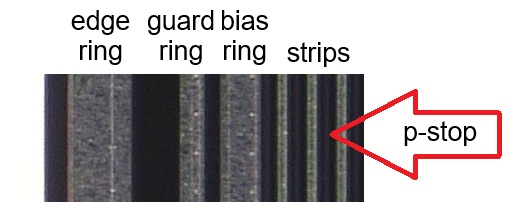}
    \caption{p-stop implant on main sensor, visible as thin traces between strips.}
    \label{fig:p-stop}
\end{figure}

Isolation between neighboring $n^+$ strips is achieved using a~narrow common p-stop implant with a surface ion concentration of approximately 4 $\times$ 10$^{12}$ cm$^{-2}$ \cite{bib:Unno-spec}. The position of the p-stop implant in the sensor is shown in Figure \ref{fig:p-stop}.

The QA test chip \cite{{bib:Orr-QA}, {bib:Ullan-QA}, {bib:Tejero-thesis}} includes, among others, an interdigitated structure with a total length identical to that of the main sensor, designed for measurement of the interstrip resistance ($R_{\textrm{int}}$) with first-neighbours.  In addition, it incorporates punch-through protection (PTP) structures to evaluate the effectiveness of the punch-through mechanism. The PTP layout comprises ten channel ends located at the bias resistor side, with the entire structure enclosed by a bias ring and a guard ring.

A subset of samples were irradiated by neutrons and gamma rays. The neutron irradiation was performed at the Jožef Stefan Institute in Ljubljana (JSI TRIGA reactor) \cite{bib:JSI}. The accuracy of neutron fluence is 10\%. The source of uncertainty is the precision of the dosimetry and errors in timing of irradiation. $^{60}$Co gamma irradiation was conducted at UJP PRAHA a.s. (Terabalt $^{60}$Co source) \cite{bib:UJP}. The sensors were maintained at room temperature during gamma irradiation. The dose rate was 9.87, 13 and 16.78 krad/min. The dose rate uncertainty is 10\%.

The irradiation proceeded in two stages:
\begin{itemize}
\item  \textit{Mid-fluence} corresponding to $21-25\,\mathrm{Mrad} + 5.1 \times 10^{14}$ $n_{\mathrm{eq}}/\mathrm{cm}^2$.
\item \textit{Full-fluence} corresponding to $66\, \mathrm{Mrad} + 1.6 \times 10^{15} n_{\mathrm{eq}}/\mathrm{cm}^2$.
\item During the irradiation campaign, several samples were accidentally neutron-irradiated up to a dose of $3.5 \times 10^{15}\,n_{\mathrm{eq}}/\mathrm{cm}^2$. These samples are named as \textit{over-irradiated} ($66~\mathrm{Mrad} + 3.5 \times 10^{15}$ $n_{\mathrm{eq}}/\mathrm{cm}^2$) and were included in tests for comparison.
\end{itemize}
   

\section{Experimental methods and results}

\subsection{Current-voltage \& capacitance-voltage characteristics}

Current-voltage ($I-V$) characteristics were measured before and after irradiation to determine if the newly implemented \mbox{p-stop} isolation fabrication processes could lead to early breakdown.

Before irradiation, 48 main sensor and 16 mini sensors from both production batches, covering all four p-stop process variations, were measured, with the mini sensors taken from the LEFT and DOWN half-moon sections of the wafers.

All sensors passed the QC specifications, with leakage current $I < 0.1\, \mu $A$/$cm$^2$ and breakdown voltages ($V_{\textrm{BD}}$) exceeding 500~V \cite{bib:Unno-spec}. No significant differences were observed between mini sensors from different half-moon positions, and no early breakdown was detected for any of the tested devices. This indicates that the modified p-stop layouts do not adversely affect the breakdown behaviour prior to irradiation.

In addition, long term stability (LTS) tests were performed before irradiation on 42 main sensors to evaluate the stability of the leakage current and stability of the breakdown voltage in $I-V$ characteristics after the LTS test. During the LTS tests, the sensors were biased at 500~V for 48 hours. All main sensors passed the stability criteria (current change <15\% during the first 24 hours \cite{bib:Unno-spec}), and no degradation of the breakdown voltage  following the LTS procedure was observed.

After irradiation to both mid- and full-fluence levels, the $I-V$ characteristics on six main sensors of p-stop types $2-4$ were measured. Prior to the measurements, all sensors were isothermally annealed for 80 minutes at 60 $^{\circ}\mathrm{C}$, and the $I-V$ measurements were performed at $-30$ $^{\circ}\mathrm{C}$, corresponding to the operational temperature of the sensors in the ATLAS experiment.
\begin{figure}[h]
    \centering
    \includegraphics[width=8.9cm]{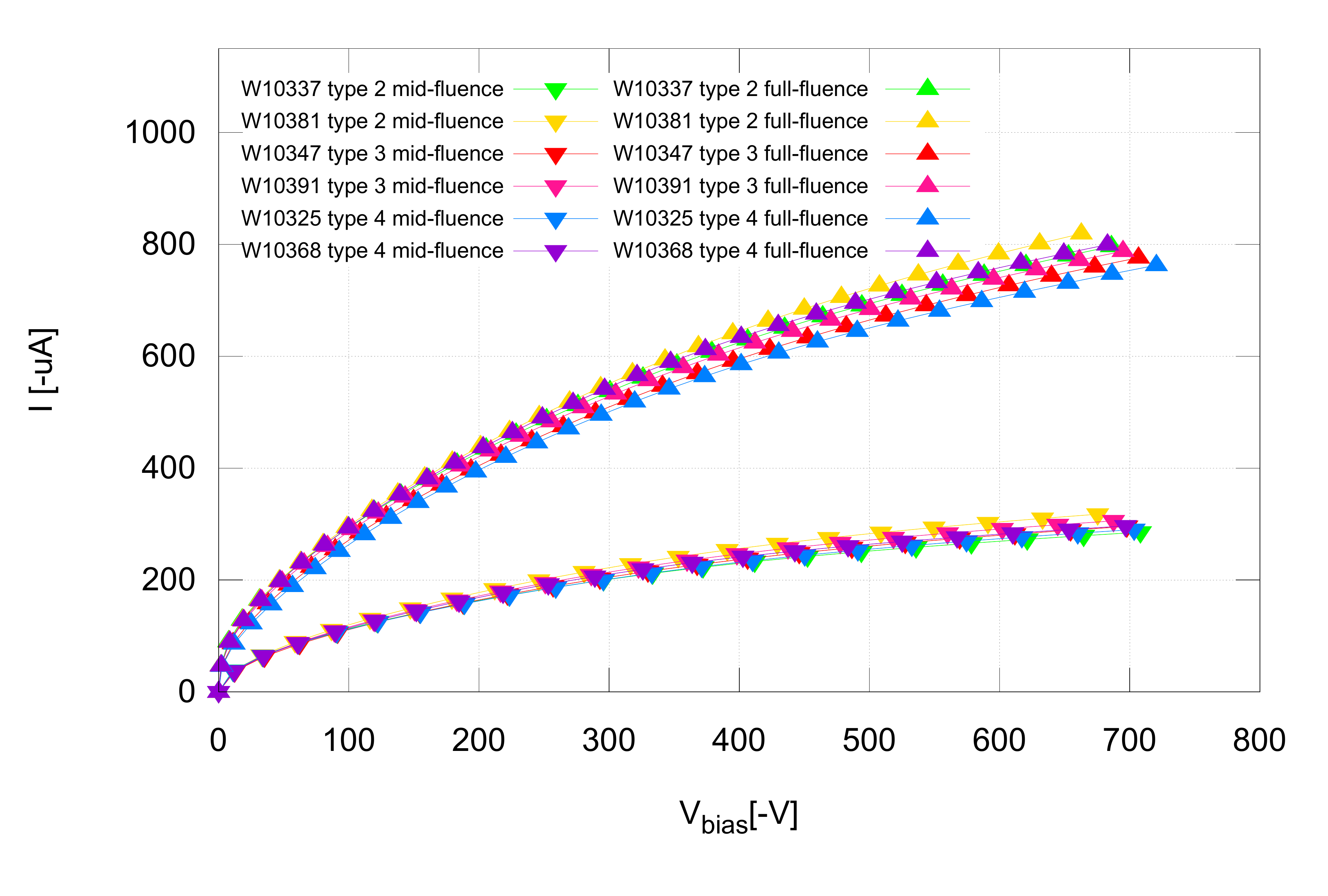}
    \caption{$I-V$ characteristics of main sensors corresponding to p-stop process types $2-4$ from both production batches, irradiated to mid- and full-fluence levels. The measurements were performed at $-30$ $^{\circ}\mathrm{C}$ after isothermal annealing for 80 minutes at 60 $^{\circ}\mathrm{C}$.}
    \label{fig:IV-after}
\end{figure}

The results are shown in Figure \ref{fig:IV-after}. No breakdown was observed up to the maximum applied voltage of 700~V, demonstrating no observable dependence of the breakdown behavior on the p-stop process variation after irradiation. All sensors fulfilled the quality requirements, with leakage currents below 2 $\mathrm{\mu}$A/cm$^2$ and breakdown voltages exceeding 500~V.

Capacitance–voltage ($C-V$) characteristics were measured on 18 unirradiated main sensors and on one main sensor irradiated to mid-fluence. All tested sensors fulfilled the QC specifications before irradiation, with full depletion voltages $V_{\textrm{FD}}$~<~350~V \cite{bib:Unno-spec}. After mid-fluence irradiation, the $V_{\textrm{FD}}$ increased to approximately 670~V.


\subsection{Punch-through voltage}
AC-coupled strip sensors can experience large voltage differences between the readout metal and implant strips, for example due to charge accumulation from beam splash accidents~\cite{bib:Hara}. To prevent capacitor breakdown, a punch-through protection (PTP) structure is implemented at the implant strip ends, providing a low-resistance current path parallel to the bias resistor once the PTP onset voltage is exceeded.

The PTP performance was evaluated using the DC method \cite{bib:Hara}. A test voltage $V_{\mathrm{test}}$ was applied between the implant and the grounded bias rail while measuring the current $I_{\mathrm{test}}$. The voltage range was $0-30$~V for unirradiated samples and $0-60$~V for irradiated samples. For irradiated sensors, $V_{\mathrm{test}}$ was applied simultaneously to two neighbouring strips to suppress interstrip effects~\cite{bib:mikestikova2020}.

The effective resistance was calculated as $R_{\mathrm{eff}} = V_{\mathrm{test}}/I_{\mathrm{test}}$. $R_{\mathrm{eff}}$ represents the parallel combination of the punch-through resistance $R_{\mathrm{PT}}$ and the polysilicon bias resistor $R_\mathrm{bias}$. The punch-through voltage $V_\mathrm{PT}$ is defined as the voltage at which $R_{\mathrm{eff}}=R_\mathrm{bias}/2$ \cite{bib:mikestikova2020}.

\begin{figure}[h!]
    \centering
    \includegraphics[width=8.4cm]{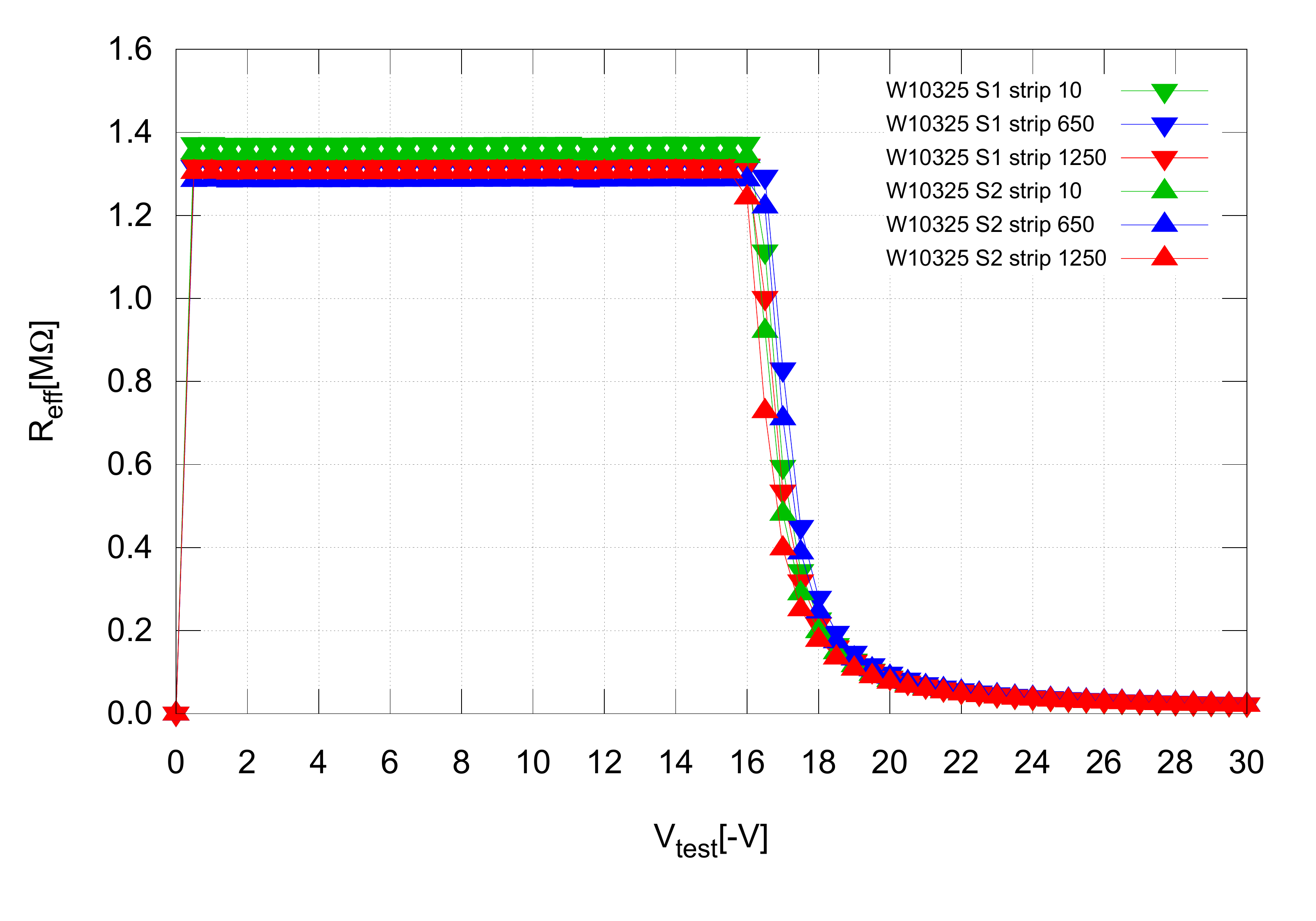}
    \caption{Effective resistance $R_\mathrm{eff}$ as a function of the applied test voltage $V_\mathrm{test}$ for a main sensor from batch VPA50958 with p-stop process type 4, measured at a bias voltage of $300$~V.}
    \label{fig:PTP-type4}
\end{figure}

Figure \ref{fig:PTP-type4} shows an example of a PTP measurement performed on a main sensor before irradiation fabricated with p-stop process type 4. The effective resistance $R_\mathrm{eff}$ is plotted as a function of the test voltage  for six different strip locations, corresponding to the left (strip 10), center (strip 650), and right (strip 1250) regions of the upper (S1) and lower (S0) sensor segments. Measurements on unirradiated samples were carried out at $20\,^\circ\mathrm{C}$ with the samples biased well above the $V_\mathrm{FD}$ ($\sim 280$ V). All tested unirradiated samples, for all four p-stop process variants, fulfil the quality control specifications for standard ATLAS18 sensors \cite{bib:Unno-spec}, with measured punch-through voltages  in the range of $12-18$~V and good uniformity across the sensor area. Type 4 results show smallest differences in the p-stop across the wafers, compared to the rest of types.

To verify the radiation hardness of the new p-stop thermal conditioning, $V_\mathrm{PT}$ was also measured on irradiated samples. Post-irradiation measurements were performed at temperatures of $-20\,^\circ\mathrm{C}$ for test chips and $-30\,^\circ\mathrm{C}$ for main sensors\footnote{The lower operating temperature used for the main sensors compensates for their higher leakage current due to the larger active volume. The measurement setup includes a 1 M$\Omega$ safety resistor, which introduces a voltage drop that is taken into account in the analysis.}.

After each irradiation step (to mid- and full-fluence), the sensors were annealed for 80 min at 60 $^{\circ}\mathrm{C}$. The bias voltages applied during the tests were $V_{\textrm{bias}} = (443 \pm 9)$~V and $V_{\textrm{bias}} = (384 \pm 13)$~V for the mid- and full-fluence sensors, respectively.

\begin{figure}[ht]
    \centering
    \includegraphics[width=8.4cm]{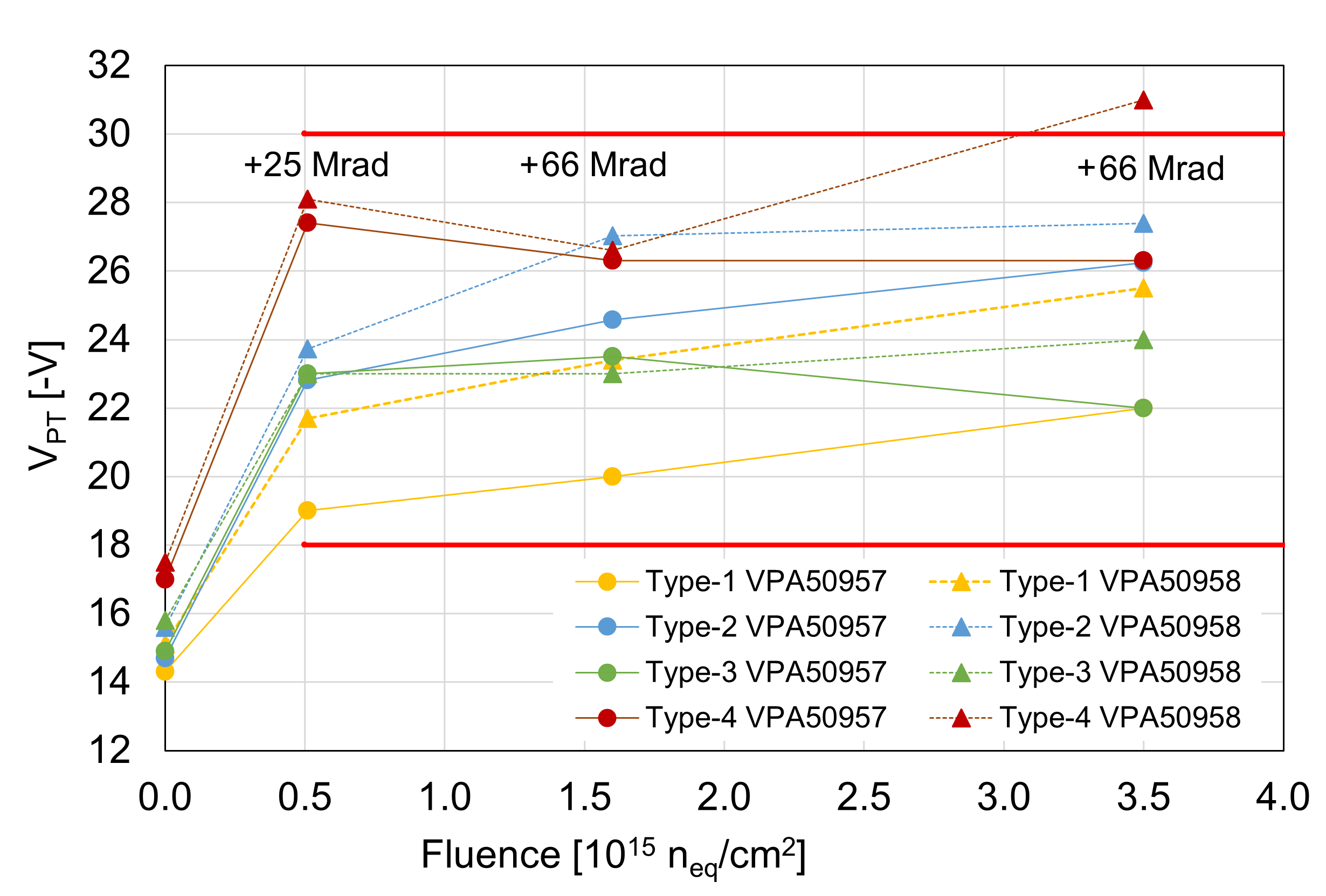}
    \caption{Dependence of punch-through voltage $V_{\textrm{PT}}$ on fluence and TID, measured on the test chips. Measurements were performed before irradiation, after mid-fluence irradiation, full-fluence irradiation, and over-irradiation. The horizontal red lines show the specification limits of irradiated test chips.}
    \label{fig:V_PT-testchips}
\end{figure}

Figure \ref{fig:V_PT-testchips} presents the results of the punch-through voltage measurements on the test chips before irradiation, after irradiation to mid- and full-fluence, and for over-irradiated samples. The results are shown for all p-stop process variants and for both production batches. 
All samples fulfil the ATLAS specifications in both irradiation states: before irradiation, the measured punch-through voltages lie within the specification range for non-irradiated sensors ($12-18$~V), while after irradiation $V_\mathrm{PT}$ increases as expected and remains within the ATLAS limits defined for irradiated sensors ($18-30$~V) \cite{bib:Unno-spec}. One over-irradiated sample slightly exceeds the limit; however, the ATLAS specification applies only up to the full-fluence and corresponding TID value.

For type 1, type 2, and type 3 samples, the $V_{\textrm{PT}}$ value increases with increasing TID and fluence. However, after the full-fluence irradiation, the type 4 samples show lower $V_\mathrm{PT}$ values compared to those measured after the mid-fluence irradiation.

\begin{figure}[ht]
    \centering
    \includegraphics[width=8.9cm]{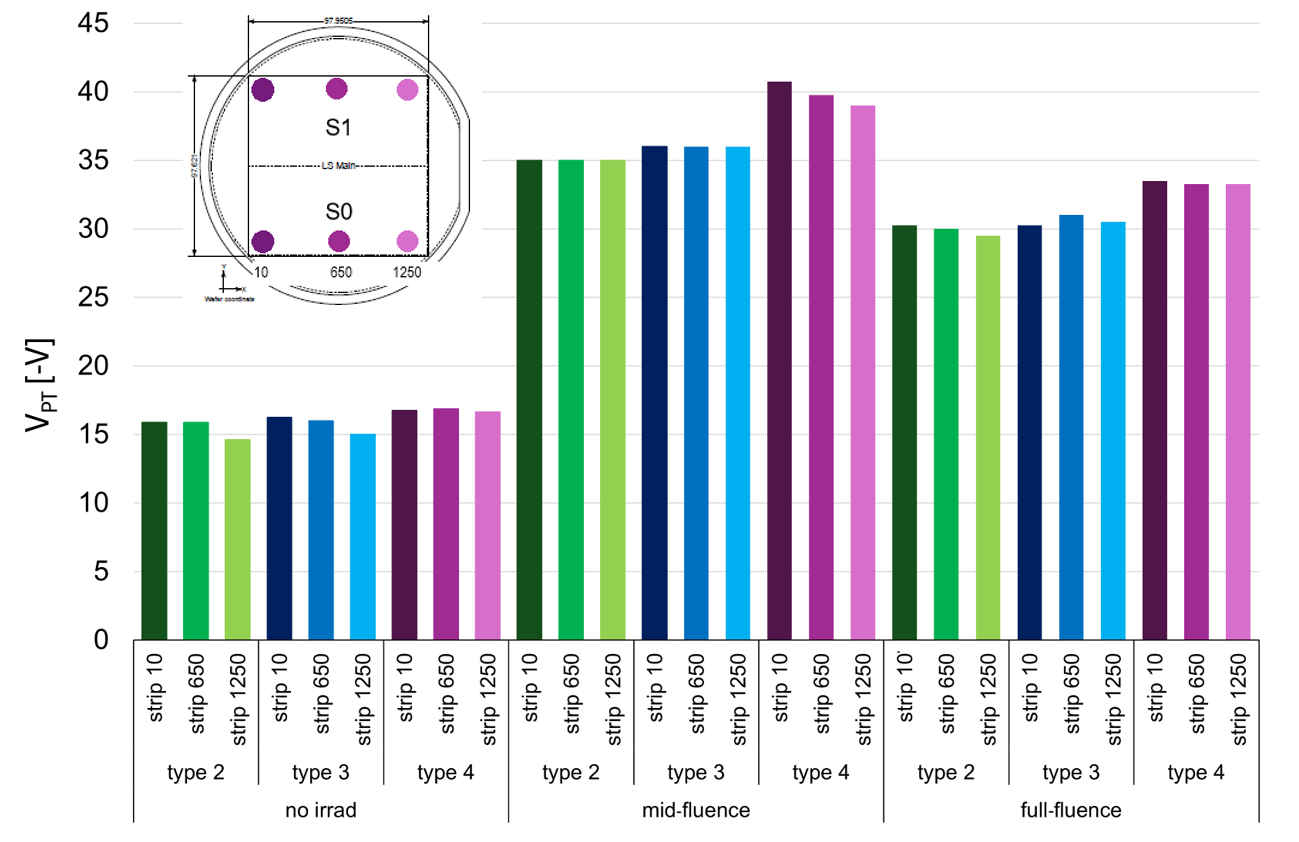}
    \caption{$V_{\textrm{PT}}$ values measured for the main sensors fabricated with p-stop process types 2, 3, and 4 before irradiation and after mid- and full-fluence irradiation. Measurements were performed at 3 different positions across the sensor area (left - strip 10, center - strip 650, and right - strip 1250). For each process type and irradiation condition, the $V_{\textrm{PT}}$ values are averaged over two production batches and the two strip segments (S0 and S1).}
    \label{fig:V_PT-main}
\end{figure}

The results of $V_\mathrm{PT}$ measurements on main sensors are shown in Figure \ref{fig:V_PT-main}. As for the unirradiated sensors, $V_\mathrm{PT}$ was measured at different positions across the sensor area (left, center, and right regions of the upper and lower segments) to assess spatial uniformity. The measurements were performed on sensors fabricated with p-stop process types 2, 3, and 4 for both batches. For each process type, only the $V_{\textrm{PT}}$ value averaged over the two batches and over two segments is shown. Results are shown for sensors irradiated to mid- and full-fluence, and, for comparison, the corresponding pre-irradiation values are also included.

As observed for the test chips, the $V_\mathrm{PT}$ of the main sensors also increases after irradiation. An interesting observation is that after mid-fluence irradiation, the $V_\mathrm{PT}$ values are higher than those measured after full-fluence irradiation. This behaviour can be explained by the fact that $V_\mathrm{PT}$ measurements in the main sensors are affected by the high leakage current caused by irradiation. The leakage current increases significantly with fluence, and the measurement setup includes a 1 M$\Omega$ safety resistor, which introduces an additional voltage drop. As a result, even when a bias voltage of 1000 V is applied, the actual voltage on the sensor is only $(384 \pm 13)$~V. This value is significantly lower than the full depletion voltage, which is expected to be around 1000 V at the full-fluence. During $V_\mathrm{PT}$ measurements of the mid-fluence sensors, the effective bias voltage was approximately 450 V, which is closer to the full depletion voltage of about 700 V. This is in agreement with results from dedicated studies that demonstrate that the $V_\mathrm{PT}$ increases with the applied bias voltage.

The largest increase in $V_\mathrm{PT}$ with irradiation is detected for samples produced with process type 4. The same behaviour for~process type 4 was also observed in the corresponding test chips (see Figure \ref{fig:V_PT-testchips}).


\subsection{Interstrip resistance}

The interstrip resistance $R_\mathrm{int}$ in a strip sensor is defined as the resistance between neighbouring $n^+$ implant strips. If the resistance between adjacent strips is too low, the signal can spread to neighbouring channels (charge sharing), resulting in degradation of spatial resolution. A sufficiently high interstrip resistance ensures good strip isolation, thereby preserving spatial resolution and maintaining an adequate signal level \cite{bib:Unno2013}.

The interstrip resistance was measured on the test chips for all process variations and for both production batches. In addition, measurements were performed on the main sensors at the same positions as for the $V_\mathrm{PT}$ measurements, to verify the uniformity across the wafer.

In the test chips, a dedicated interdigitated strip structure is implemented. Two groups of strips are connected in parallel, with strips from the two groups alternating while maintaining the nominal sensor strip pitch. Two types of interdigitated structures are available, corresponding to strip lengths of 2.4 cm and 4.8 cm. The interstrip resistance was determined by measuring the $I-V$ characteristics between the two strip groups.

For the main sensors, the interstrip resistance was measured by contacting three adjacent implant DC pads with probe needles. A voltage $V_\mathrm{app}$ was applied to the central strip, and the interstrip current $I_\mathrm{int}$ was measured on the two neighbouring strips. The interstrip resistance was calculated as

$$R_\mathrm{int}=\frac{2}{\mathrm{d}I_\mathrm{int}/\mathrm{d}V_\mathrm{app}}.$$

Measurements of irradiated samples were performed after annealing for 80 minutes at 60 $^{\circ}\mathrm{C}$ and at cold temperatures of $-20, -25,$ or $-30$ $^{\circ}\mathrm{C}$, depending on the ATLAS ITk institution performing the measurement. For unirradiated sensors, the interstrip resistance is of the order of several tens of G$\Omega$.

According to the ATLAS specifications \cite{bib:Unno-spec}, the interstrip resistance must be at least ten times larger than the bias resistance $R_\mathrm{bias}$. This corresponds to $R_\mathrm{int} \geq 18\,\mathrm{M}\Omega$ at $-20$ $^{\circ}\mathrm{C}$ for an applied bias voltage of 500 V. 

\begin{figure}[h!]
    \centering
    \includegraphics[width=8.6cm]{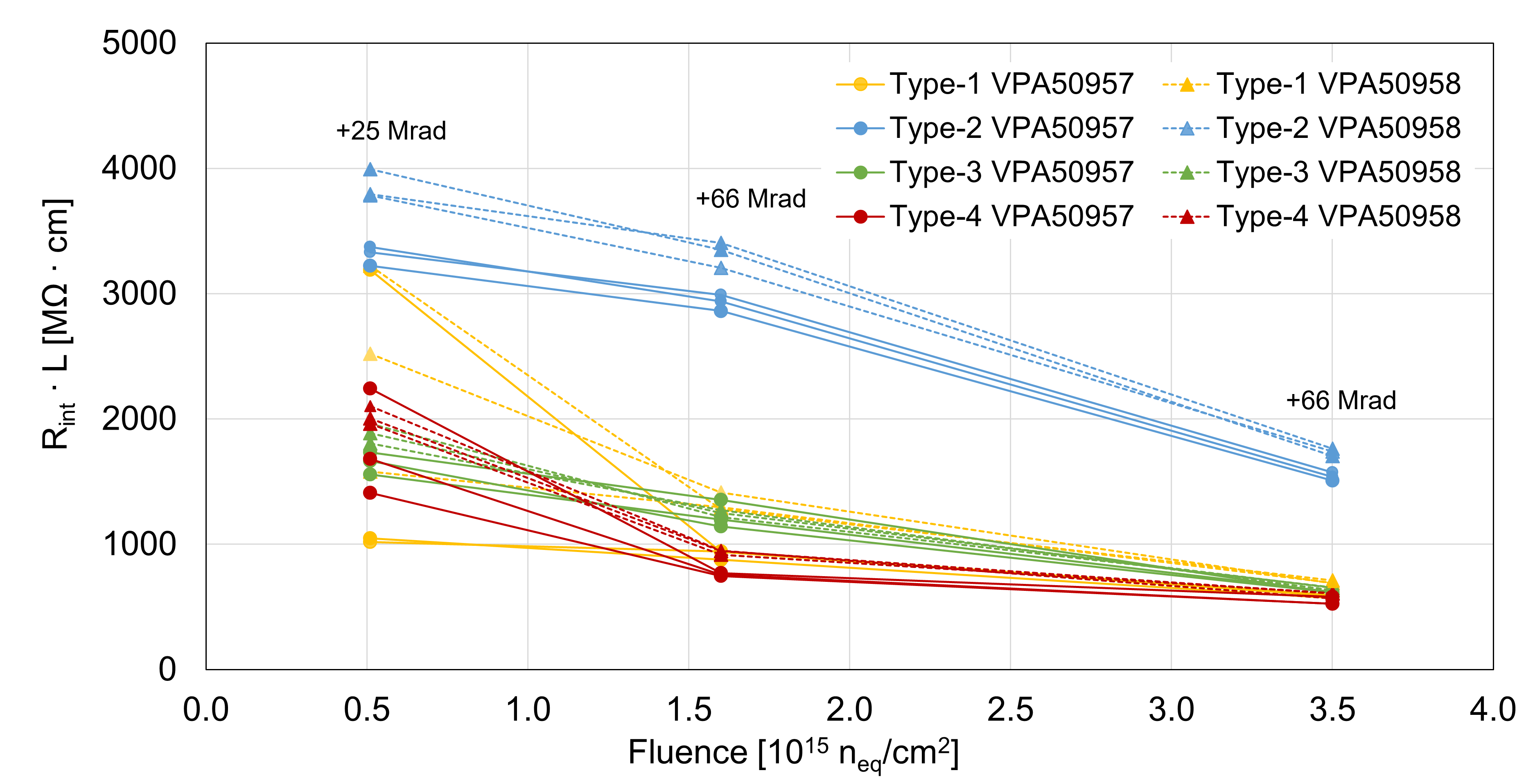}
    \caption{Interstrip resistance measured on interdigitated structures in test chips after mid- and full-fluence irradiation, and over-irradiated samples. The values are normalized to $-20$ $^{\circ}\mathrm{C}$ and to a strip length of 1 cm.}
    \label{fig:Rint_testchips}
\end{figure}

Figure \ref{fig:Rint_testchips} shows the $R_\mathrm{int}$ values measured on the test chips after mid- and full-fluence irradiation, as well as on over-irradiated samples for comparison. Since the measurements were performed at different temperatures in various institutions and the interstrip resistance is temperature dependent all values are normalized to $-20$ $^{\circ}\mathrm{C}$ \cite{{bib:Unno2013}, {bib:mikestikova2016}}. In addition, the interstrip resistance values are normalized to a strip length of 1 cm.

The strongest degradation of interstrip isolation is observed for the samples fabricated with process type 4, while the sensors produced with process type 2 exhibit the highest interstrip resistance. All measured values of $R_\mathrm{int}$ are well above the ATLAS specifications.

\begin{figure}[ht]
    \centering
    \includegraphics[width=8.5cm]{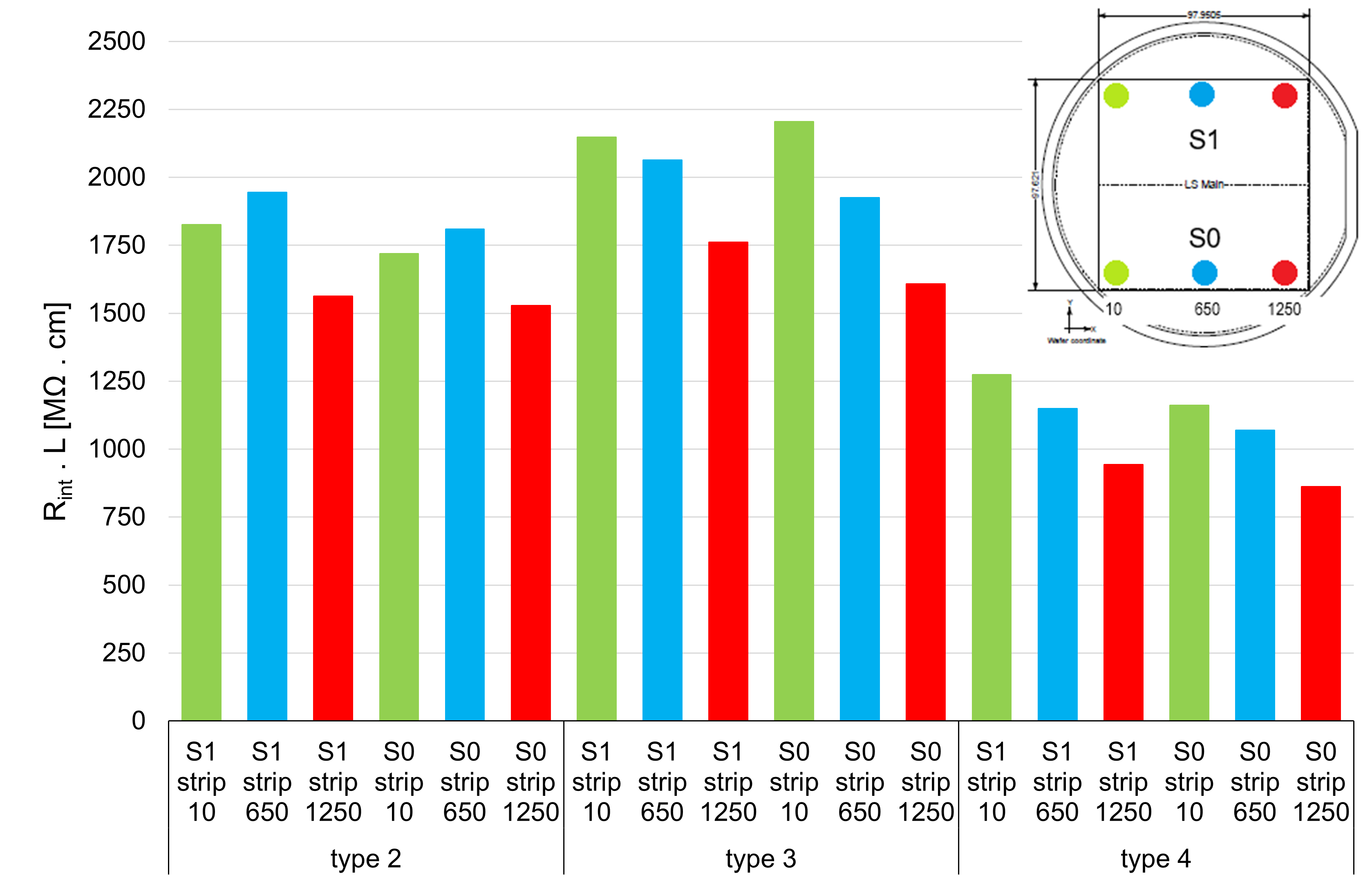}
    \caption{Interstrip resistance measured on the main sensors fabricated with process types 2, 3, and 4 after full-fluence irradiation. The measurements were performed across the sensor (strip 10 - left, strip 650 - center and strip 1250 - right) and for both strip segments (lower S0 and upper S1). The values are normalized to $-20$ $^{\circ}\mathrm{C}$, to a strip length of 1 cm. $V_{\textrm{bias}} \approx $ $380$~V.}
    \label{fig:Rint_main}
\end{figure}

The interstrip resistance measured on the main sensors fabricated with process types 2, 3, and 4 and irradiated to full-fluence is presented in Figure \ref{fig:Rint_main}. The $R_\mathrm{int}$ values were taken at $V_{\textrm{bias}} = (384 \pm 13)$~V and are normalized to $-20$ $^{\circ}\mathrm{C}$.

The main sensors fabricated with process type 4 exhibit slightly lower $R_\mathrm{int}$ values compared to process types 2 and 3, in agreement with the results obtained from the test chips. The large interstrip resistance observed in the type 2 test chips is not confirmed in the type 2 main sensors. The $R_\mathrm{int}$ values show some variation across the wafer for all process types. The left side of the sensor (lower strip numbers) exhibits slightly higher interstrip resistance than the right side (higher strip numbers), with the largest difference observed for process type 3.

In type 4 samples, the observation of the lower $R_\mathrm{int}$ (Figure \ref{fig:Rint_main}) and the higher $V_\mathrm{PT}$ (Figure \ref{fig:V_PT-main}) contradicts the expectation based on the underlying p-stop density: the higher the density, the larger the $R_\mathrm{int}$ and the $V_\mathrm{PT}$ \cite{{bib:Lindgren},{bib:Unno-HSTD14}}. This discrepancy is currently under investigation.

\subsection{Strip tests}

Low strip isolation area can also be identified by performing the \textit{full strip test}, which is routinely used by ATLAS ITk institutes for quality control testing of production sensors \cite{bib:Federicova}. This test assesses the quality of the manufacturing process and the uniformity of the electrical properties across the entire wafer. In this procedure, each individual strip of the sensor is contacted and its impedance to ground is measured. This allows the identification of potential defects such as metal shorts, broken implants, faulty bias resistors, low interstrip isolation, as well as pinholes or punch-throughs in the dielectric layers.

A loss of interstrip isolation indicates the presence of unintended electrical connections between adjacent strips or malfunctioning bias resistors. In practice, this occurs when the interstrip resistance becomes comparable in magnitude to the bias resistance $R_\mathrm{bias}$, leading to degraded strip isolation. 

The temperature and relative humidity during the measurements were monitored and controlled to ensure stable conditions. During the full strip test, the sensor is biased at a voltage sufficiently high to achieve full strip isolation; a bias voltage of $150$~V is typically applied.

Before irradiation, a total of 26 main sensors were subjected to the full strip test. The sample included sensors of all four process types from both production batches. Overall, no abnormal behavior was observed. For all sensor types, the bias resistance $R_\mathrm{bias}$ and the coupling capacitance $C_\mathrm{coupl}$ were within the ATLAS specification limits.

A mild decrease of the coupling capacitance (below 3\%) and of $R_\mathrm{bias}$ (below 9\%) was observed across the strip segments. This behaviour is consistent with trends seen in standard QC full strip tests during production \cite{bib:Federicova}. No correlation was found between the measured $V_\mathrm{PT}$ and the parameters extracted from the full strip test. 

\begin{figure}[h]
    \centering
    \includegraphics[width=8.3cm]{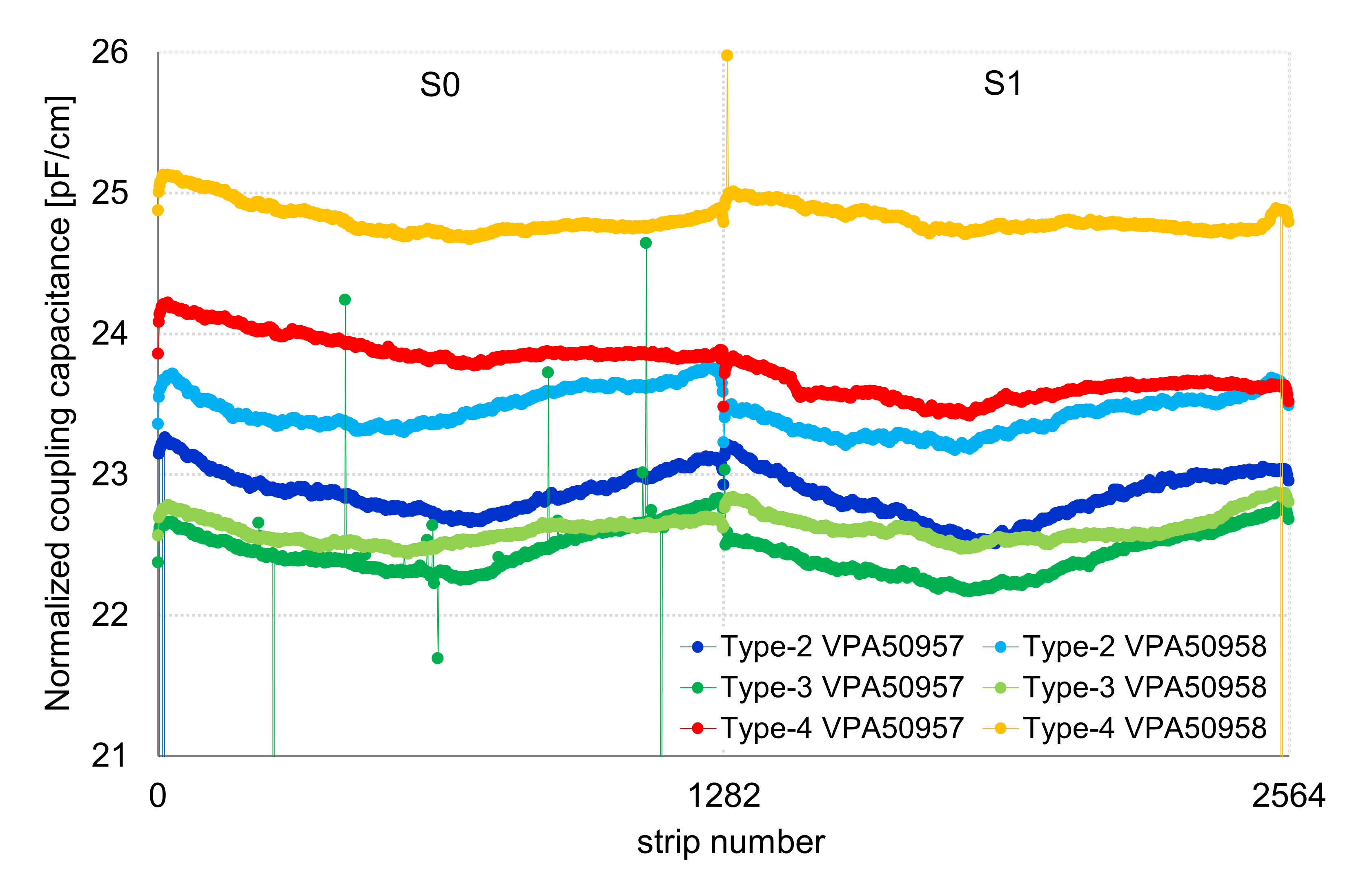}
    \caption{An example of results of full strip test for the $C_\mathrm{coupl}$ measured on both strip segments (S0, S1: strip numbers are assigned sequentially from S0 through S1) after mid-fluence irradiation at $-20\,^{\circ}\mathrm{C}$. Sensors were biased to $90$~V during tests.}
    \label{fig:striptest-mid}
\end{figure}

After irradiation to mid-fluence, six sensors of process types 2, 3, and 4 from both production batches were measured using the full strip test. The measurements were performed at a temperature of $-20\,^{\circ}\mathrm{C}$. The results of the strip test for the coupling capacitance are shown in Figure \ref{fig:striptest-mid}. All tested sensors fulfilled the quality control requirements of the ATLAS ITk strip test.

Localized peaks observed in the coupling capacitance of type~3 sensor from batch VPA50957 are attributed to contact issues during the measurement of segment S0 rather than to sensor-related effects. The strip test was performed at a reduced voltage. This corresponds to applying $500$~V through a 1~M$\Omega$ safety resistor, which results in an effective bias voltage of approximately $90$~V at $-20\,^{\circ}\mathrm{C}$ \footnote{See Figure~\ref{fig:IV-after} for the current that was measured at $-30\,^{\circ}\mathrm{C}$.}, instead of the standard $150$~V.

\begin{figure}[h!]
    \centering
    \includegraphics[width=8.3cm]{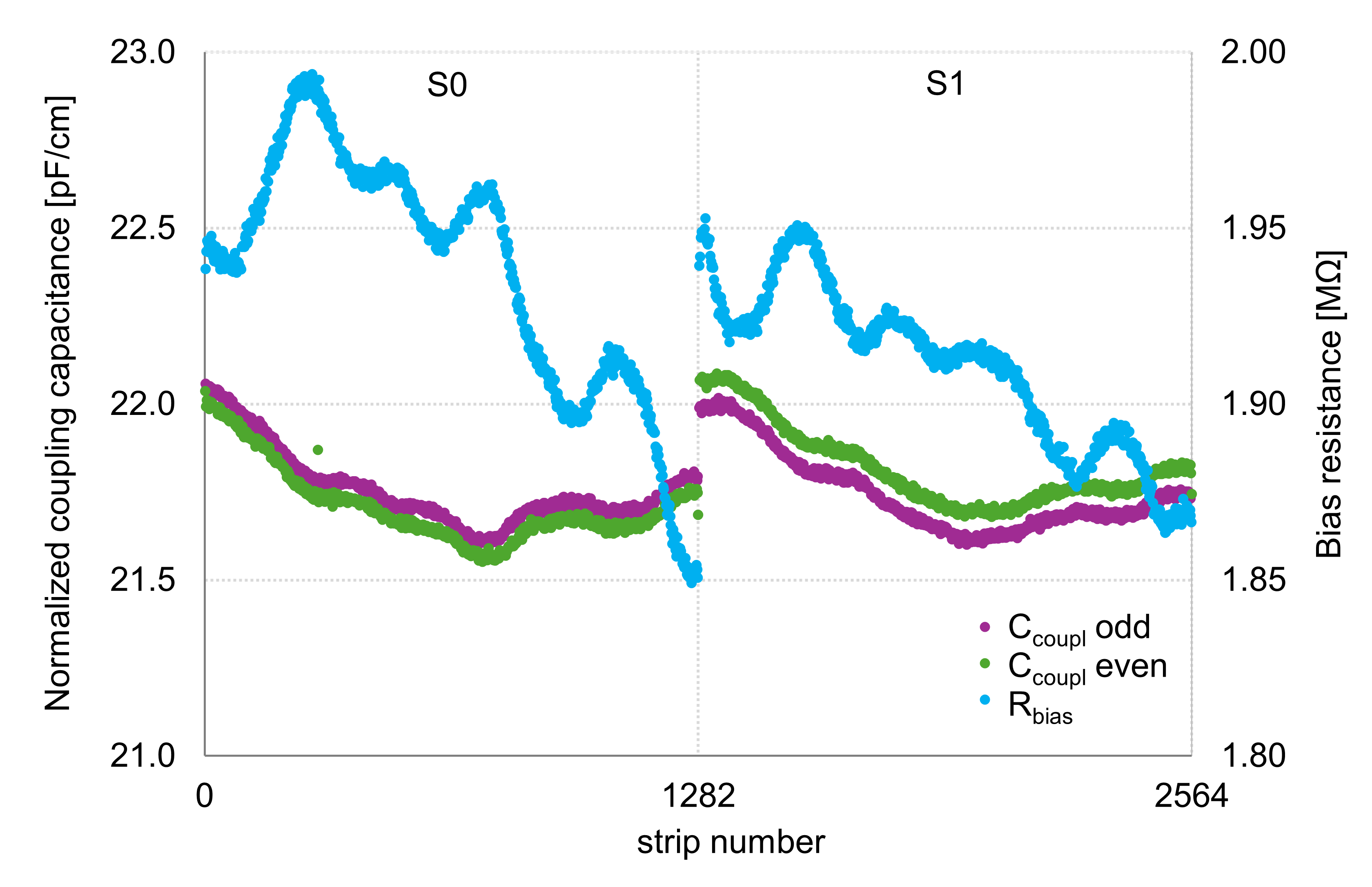}
    \caption{The full strip test results ($C_\mathrm{coupl}$ and $R_\mathrm{bias}$) measured on both strip segments (S0, S1: strip numbers are assigned sequentially from S0 through S1) of type 3 sensor from batch VPA50957 after full-fluence irradiation at $-30\,^{\circ}\mathrm{C}$. Sensor was biased to $256$~V during the test.}
    \label{fig:striptest-full}
\end{figure}

After full-fluence irradiation, six sensors of process types 2, 3, and 4 were tested. These measurements were carried out at a temperature of $-25\,^{\circ}\mathrm{C}$ and $-30\,^{\circ}\mathrm{C}$ in order to reduce the leakage current and thus minimise the voltage drop across the safety resistor. All six sensors showed normal behaviour and passed the \mbox{ATLAS} quality control specifications~\cite{bib:Unno-spec}. Figure \ref{fig:striptest-full} presents  the full strip test results for the coupling capacitance and bias resistance of type~3 sensor from batch VPA50957 after full-fluence irradiation. In contrast to the mid-fluence measurement, no contact-related issues were observed.

In summary, the full strip tests performed before irradiation and after irradiation to both mid- and full-fluence levels did not reveal any problematic strips for any of the investigated p-stop process variations. No degradation of strip isolation was observed in any region of the sensors, demonstrating stable and uniform strip behavior across all tested irradiation states.

\subsection{Charge collection efficiency}

The charge collection efficiency (CCE) quantifies the fraction of charge generated in the silicon bulk by an ionizing particle that is collected at the readout electrodes and is a key indicator of detector performance after irradiation. To investigate the impact of different p-stop process types on CCE, measurements were performed on irradiated mini sensors using $^{90}$Sr $\beta$-rays, with the resulting ionisation pulse height read out by an Alibava system \cite{{bib:Orr-QA},{bib:Hernandez}}.

\begin{figure}[h!]
    \centering
    \includegraphics[width=8.6cm]{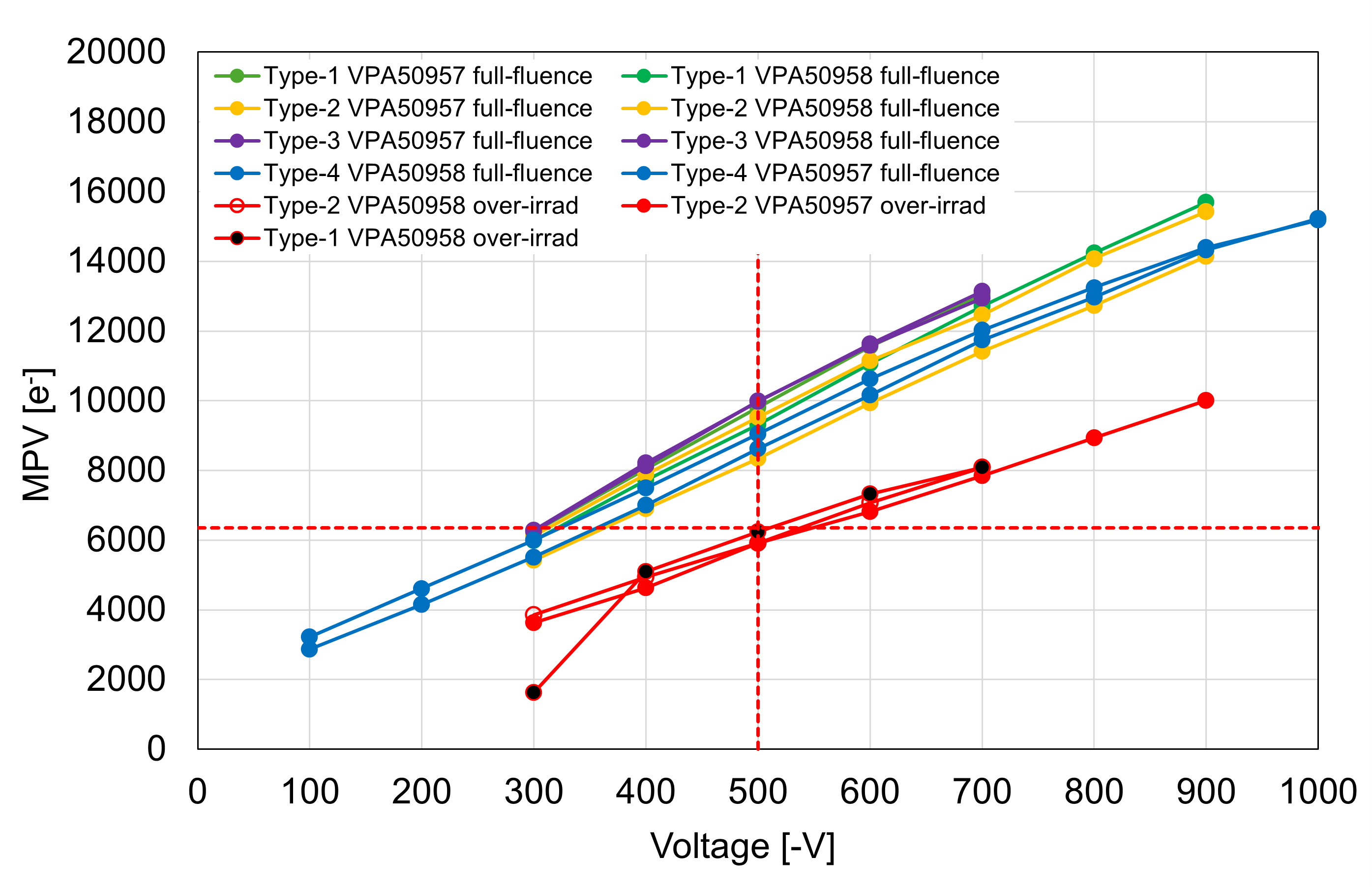}
    \caption{Charge collection efficiency of mini sensors. Red dashed lines show the specification threshold 6350 $e^-$ at $500$~V.}
    \label{fig:CCE}
\end{figure}

CCE measurements were carried out on mini sensors of all four p-stop process types irradiated to full fluence, as well as on over-irradiated samples. The results are shown in Figure~\ref{fig:CCE}. All sensors irradiated to full fluence meet the specification threshold for standard ATLAS18 sensors (6350 $e^-$ at 500~V \cite{bib:Unno-spec}). In contrast, the over-irradiated samples exhibit reduced CCE values and do not fulfil the specification. No degradation of CCE attributable to the p-stop process type was observed.

\section{Conclusions}

The goal of this evaluation program was to assess the impact of fabrication process variations on interstrip isolation performance and to provide feedback to HPK for future sensor development. Overall, the results demonstrate good sensor quality for all evaluated p-stop types, with no indication of early breakdown in $I-V$ measurements. Although small differences in punch-through protection behaviour were observed, all samples fulfilled the ATLAS ITk QC specifications both before and after irradiation.

Sensors fabricated with p-stop process type 4 exhibited the most uniform punch-through voltage distribution before irradiation. Interstrip resistance decreased with increasing radiation dose and fluence but remained above the specification limits for all process types, with type 2 showing the highest post-irradiation $R_\mathrm{int}$. The observed differences in $V_\mathrm{PT}$ and $R_\mathrm{int}$ among the p-stop implant variants are small and negligible with respect to ITk strip sensor specifications and the expected operating conditions.

Full strip tests before and after full-fluence irradiation showed no problematic regions with degraded interstrip isolation for any of the investigated process types.

These results confirm the reliability of the evaluated p-stop designs and indicate specific advantages for future sensor optimization.


\section*{Acknowledgements}

This work was supported by the European Structural and Investment Funds and the Ministry of Education, Youth and Sports of the Czech Republic via projects LM2023040 CERN-CZ, and FORTE CZ.02.01.01/00/22\_008/0004632, the Canada Foundation for Innovation under project number 36248; additional resources were provided by the Natural Sciences and Engineering Research Council of Canada. The authors would like to thank the crew at the TRIGA reactor in Ljubljana for help with irradiations. The authors acknowledge the financial support from the Slovenian Research and Innovation Agency (research core funding No. P1-0135 and project No. J1-3032). This work is part of the Spanish R\&D grant PID2021-126327OB-C22, funded by MICIU/ AEI/10.13039/501100011033 and by ERDF/EU, R\&D grant PID2021-126327OB-C21, funded by MCIN/ AEI/10.13039/501100011033 / FEDER, UE. This work at QMUL was supported by STFC grants ST/W000474/1, ST/S00095X/1, ST/X001431/1, ST/R00241X/1. The work at SCIPP was supported by the US Department of Energy, grant DE-SC0010107.

Copyright 2026 CERN for the benefit of the ATLAS Collaboration. Reproduction of this article or parts is allowed as specified in the CC-BY-4.0 license.


\end{document}